\documentclass[aps,prb,showpacs,twocolumn,bibnotes,nofootinbib,superscriptaddress]{revtex4-1}

\usepackage{amssymb}
\usepackage{amsbsy}
\usepackage{amsmath}
\usepackage{amsfonts}
\usepackage{mathrsfs}
\usepackage{latexsym}
\usepackage[english]{babel}
\usepackage{url}
\usepackage[]{graphicx}
\usepackage{subfigure}
\usepackage{hyperref}
\begin{document}

\title{Nambu monopoles interacting with lattice defects in two-dimensional artificial square spin ice}

\author{R. C. Silva}
\email{rodrigo.costa@ufv.br}
\affiliation{Departamento de F\'{i}sica, Universidade Federal de Vi\c{c}osa, 36570-000 - Vi\c{c}osa - Minas Gerais - Brazil.}

\author{R. J. C. Lopes}
\affiliation{Departamento de F\'{i}sica, Universidade Federal de Vi\c{c}osa, 36570-000 - Vi\c{c}osa - Minas Gerais - Brazil.}

\author{L. A. S. M\'{o}l}
\affiliation{Departamento de F\'{i}sica, Universidade Federal de Vi\c{c}osa, 36570-000 - Vi\c{c}osa - Minas Gerais - Brazil.}

\author{W. A. Moura-Melo}
\affiliation{Departamento de F\'{i}sica, Universidade Federal de Vi\c{c}osa, 36570-000 - Vi\c{c}osa - Minas Gerais - Brazil.}

\author{G. M. Wysin}
\affiliation{Departamento de F\'{i}sica, Universidade Federal de Vi\c{c}osa, 36570-000 - Vi\c{c}osa - Minas Gerais - Brazil.}
\affiliation{Department of Physics, Kansas State University, Manhattan, Kansas State 66506-2601, USA}

\author{A. R. Pereira}
\affiliation{Departamento de F\'{i}sica, Universidade Federal de Vi\c{c}osa, 36570-000 - Vi\c{c}osa - Minas Gerais - Brazil.}

\date{\today}

\begin{abstract}

The interactions between an excitation (similar to a pair of Nambu monopoles connected by their associated string) and a lattice defect are studied in an artificial two-dimensional square spin ice. This is done by considering a square array of islands containing only one island different from all others. This difference is incorporated in the magnetic moment (spin) of the ``imperfect"  island and several cases are studied, including the special situation in which this distinct spin is zero (vacancy). We show that the two extreme points of a defective island behave like two opposite magnetic charges. Then, the effective interaction between a pair of Nambu monopoles with the defective island is a problem involving four magnetic charges (two pairs of opposite poles) and a string. We also sketch the configuration of the field lines of these four charges to confirm this picture. The influence of the string on this interaction decays rapidly with the string distance from the defect.

\end{abstract}
\pacs{75.75.-c, 75.40.Mg, 75.50.-y, 75.30.Hx
}

\maketitle

\section{Introduction}

Artificial spin ices~\cite{Wang} are systems composed by an array of lithographically defined two-dimensional ($2d$) ferromagnetic nanostructures with single-domain islands (elongated permalloy nanoparticles, in general), where the net magnetic moment of each island is assumed to be well approximated by an Ising-like spin (for a regime out of the Ising behavior in a single elliptic island, see Ref.~\onlinecite{Wysin}). They can be produced in diverse types of geometries with lattices like the square~\cite{Wang}, brickwork~\cite{Li}, honeycomb or kagome~\cite{Ladak, Mengotti}, triangular~\cite{Mol12} etc. Recently, these artificial materials have been objects of intense experimental and theoretical investigations~\cite{Wang, Li, Ladak, Mengotti,Mol12, Moller1, Ke, Mol1, Mol2, Morgan, Moller2, Zabel, Budrikis, Reichhardt, Kapaklis, Silva} associated mainly with the appearance of collective excitations that are expected to behave like magnetic monopoles.

\begin{figure}[hbt]
		\centering
		\includegraphics[width=8.2cm]{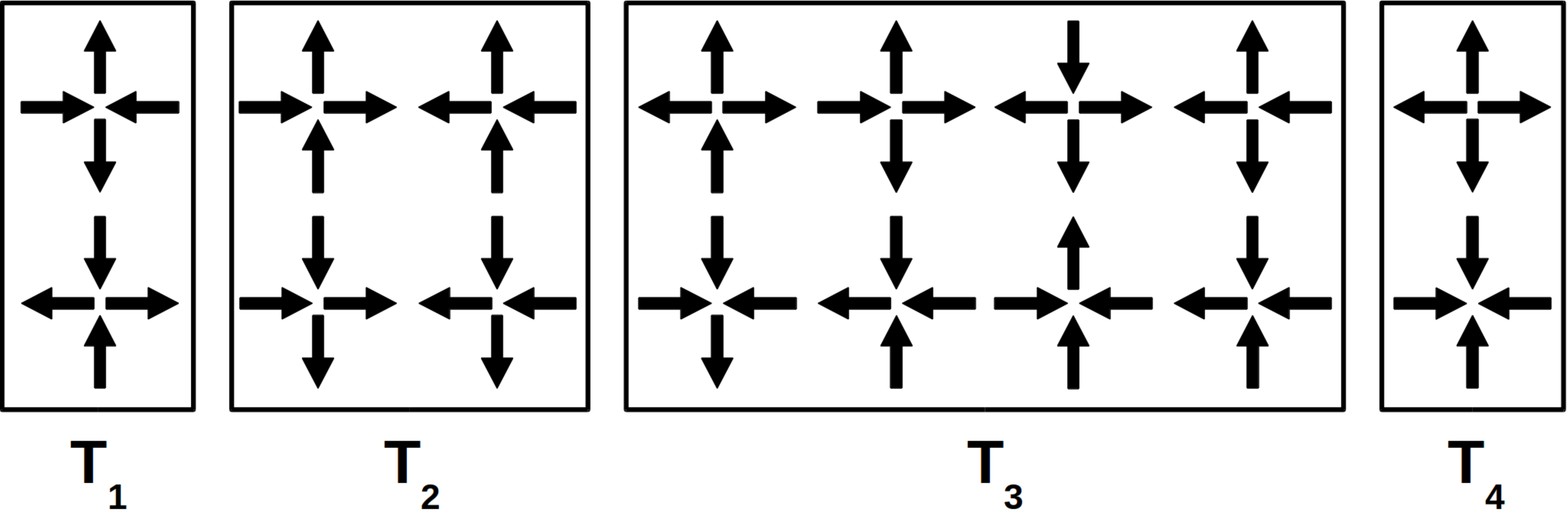}
		\caption{The four topologies for the vertices in an artificial square spin ice. The energy of these topologies increases from left to right. Topologies $T_{1}$ and $T_{2}$ obey the ice rule (two-in/two-out) but, they are not degenerate. Topology $T_{3}$ exhibits the configurations with three-in/ one-out or one-in/ three-out while in the topology $T_{4}$, one has four-in or four-out. Topology $T_{1}$ gives the ground state. Topology $T_{2}$ is associated with the string-like excitations and $T_{3,4}$ are associated with magnetic monopole-like excitations.}
		\label{fig:squareice_topology}
	\end{figure}

The theoretical and experimental studies concerning the physical properties of the ground state and excitations of the artificial square spin ices have deserved a great deal of attention in recent years~\cite{Moller1, Mol1, Mol2, Morgan, Moller2, Silva}. In this system, there are four Ising spins at each vertex and they can be distributed in sixteen configurations grouped in four different topologies (see Fig.~\ref{fig:squareice_topology}). Nowadays, it is well established that its ground state has a configuration that obeys the ice rule (two spins point inward while the other two point outward in each vertex, but following only topology $T_{1}$ as shown in Fig.~\ref{fig:squareice_topology}). In addition, theoretical results show that the elementary excitations are quasi-particles akin to opposite magnetic monopoles connected by an energetic string~\cite{Mol1, Mol2, Morgan, Moller2, Silva} (this string is an oriented line of dipoles passing by vertices that obey the ice rule, but sustaining only topology $T_{
2}$). The string energy is associated with the fact that the ice rule is not degenerate in two dimensions, since topology $T_{2}$ has more energy than topology $T_{1}$~\cite{Wang,Mol1}. These monopoles can be then referred to as Nambu monopoles due to their similarities with the monopoles studied by Nambu in the $1970$'s in a modified Dirac monopole theory~\cite{Nambu}. Indeed, in Nambu's theory~\cite{Nambu}, the string connecting the opposite monopoles has the following features~\cite{Nambu}: (a) The end points of the string (of length $Z$) behave like monopoles interacting by a Yukawa potential. The string energy is proportional to $Z$. Therefore, for a sufficiently long string, the string energy is dominant and for a short string, the Yukawa interaction becomes important. (b) The string is oriented, having an intrinsic sense of polarization, like a magnet. Therefore, these Nambu particles have a phenomenology similar to that observed~\cite{Mol1,Mol2,Mol12} for artificial square spin ices. Really, for a 
simple comparison, we notice that the forces that bind ``monopoles'' and ``anti-monopoles'' in a $2d$ artificial square spin ice are of two kinds\cite{Mol1,Mol2}. One is the tension $b$ of an energetic string; the other is the Coulomb force (which is a particular case of the Yukawa force) given by $q/R^2$, where $q$ measures the strength of the interaction and $R$ is the distance between the poles. These features show that the excitations present in the $2d$ artificial square ice are more similar to that of Nambu theory than that of Dirac theory, justifying the use of this terminology for distinguishing the ``monopole'' excitations in different spin ice materials. Then, differently from the three-dimensional crystalline spin ices~\cite{Castelnovo} in which the string is observable but does not have energy, in the $2d$ case, there is an oriented and energetic one-dimensional string of dipoles that terminates in the monopoles with opposite charges. This string costs an energy equal to $bX$, where $X$ is the 
string length. Thus, the interaction potential between two opposite charges is generally given by $V_{N}(R,X,\phi)=q(\phi)/R + b(\phi)X+c(\phi)$, where $\phi$ is the angle that the line joining the monopoles makes with the $x$-axis of the array (there is a small anisotropy in the interaction\cite{Mol2}). Numerically, the theoretical values~\cite{Mol2,Morgan} for the constants arising in the potential $V_{N}(R,X,\phi)$ are $q(0)\approx-3.88 \, Da$, $b(0) \approx 9.8 \, D/a$ while $q(\pi/4)\approx-4.1 \, Da$, $b(\pi/4) \approx 10.1 \, D/a$, where $D=\mu_{0}\mu^{2}/4 \pi a^{3}$ is the coupling constant for the dipolar interaction among the islands, $a$ is the lattice spacing, $\mu_{0}$ and $\mu$ are the vacuum magnetic permeability and the island's magnetic moment respectively. The constant $c(0) \approx 23 \, D$ is associated with the pair creation energy, $E_{c} \approx 29 \, D$~\cite{Mol2, Morgan}, which is independent of $\phi$. The modulus of the magnetic charge is, therefore, given by $|Q_{M}(\phi)|= \
sqrt{(4 \pi |q(\phi)|/\mu_{0})}$.

Although the fabrication of these systems is relatively easy, the limitations of the lithographic techniques are a significant barrier for building ``perfect'' arrays with identical islands disposed at all lattice sites. Indeed, a large number of samples are made, for example, with malformed islands, resulting in a quenched disorder in the system (see, for example, Refs.~\onlinecite{Mengotti, Kohli12, Budrikis, Budrikis12, Budrikis12b, Dauheimer11, Phatak12, Pollard12}). On the other hand, defects can also be introduced intentionally into the system (for instance, by removing an island from the array or making, by design, some islands with holes\cite{Mengotti,Rahm03,Pereira05}). Then, as it happens with natural materials, lattice defects could also play an important role in the properties of these artificial frustrated compounds. Our primary aim in this paper is to study the effects that a single defective island causes on the elementary excitations of the artificial square spin ices. Our results show that 
the defective island induces magnetic charges on adjacent vertices, giving in this way further and strong support to the magnetic monopole  picture for the excitations of the artificial square spin ice. This picture is also corroborated by the determination of the magnetic field lines produced by excitations. Our results also suggest that by changing the shape and size of some islands of the system it may be possible to tailor design systems with desired properties.

\section{DEFECTIVE ARTIFICIAL SQUARE ICE}

Defects may be either naturally present in the system (due to the limitations of experimental techniques) or intentionally introduced in the artificial arrays. For example, one could remove an island (``spin'') from a $2d$ square lattice. Thus, it is important to study the effects of these defects on the properties of the system. Here, we will consider an arrangement of dipoles similar to that accomplished in Ref.~\onlinecite{Wang}; however, at a particular site (denoted by site $l$) the island is defective and may be larger or smaller than the other ones. In our calculations, such island deformation is incorporated in its magnetic moment which is proportional to the island volume (the spin or magnetic moment $\vec{S}_{l}$ is considered to be proportional to the island's volume). In our approach, the magnetic moment of each island is replaced by a unity Ising-like point dipole at its center ($|\vec{S}_i|=1$) which is restricted to point along the $x$ or $y$ direction depending on its position for all islands,
 except for the defective one, site $l$, whose magnitude is chosen in the interval $0 \leq |\vec{S}_{l}| \leq 2$. Comparisons between an Ising and a non Ising-like description of the nanoislands can be found, for example, in Refs.~\onlinecite{Wysin, Wysin12, Phatak12}. Note that the special limiting case of a missing island ($\vec{S}_{l}=0$) is included in our range.
In this way, the system is described by the following Hamiltonian\cite{Mol1,Mol2}:
\begin{equation}\label{Hamiltonian}
	H = Da^{3} \sum_{i \neq j} \left[ \dfrac{\vec{S}_{i} \cdot \vec{S}_{j}}{|\vec{r}_{ij}|^{3}} -3 \dfrac{(\vec{S}_{i} \cdot \vec{r}_{ij})(\vec{S}_{j} \cdot \vec{r}_{ij})}{|\vec{r}_{ij}|^{5}} \right],
\end{equation}
where $\vec{r}_{ij}$ is the vector that connects sites $i$ and $j$, $D=\mu_{0}\mu^{2}/4 \pi a^{3}$ is the coupling constant for the dipolar interactions. In all calculations, periodic boundary conditions were implemented by means of the Ewald Summation\cite{ZWang, Weis}.

\begin{figure}[hbt]
		\centering
		\includegraphics[width=9.08cm]{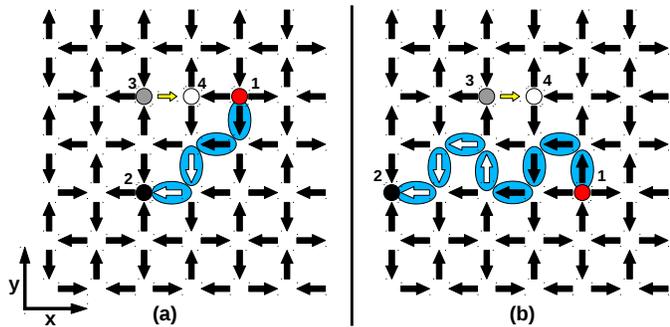}
		\caption{(Color online) Particular configuration of excitations in a lattice with a defective island (yellow (light gray) arrow between numbers $3$ and $4$). We use two basic shortest strings in the separation process of the magnetic charges, which are indicated by the marked white arrows. Pictures (a) and (b) exhibit two kinds of strings, $I$ and $II$, respectively. The black circle is the positive charge while the red (dark gray) circle is the negative charge.}
		\label{fig:defect_lattice}
	\end{figure}

In the system with a single deformed island defect, we have observed, by using a simulated annealing process (see Refs.~\onlinecite{Mol1,Mol2}) that, the ground state is the same as that of a perfect array (all vertices obeying the ice rule with topology $T_{1}$). However, at the two particular adjacent vertices shared by the defective island, there is a nonzero net magnetic moment due to the unbalance caused by the defect, since its spin is smaller (or greater) than the other three normal spins that complete the vertex (see Fig.~\ref{fig:defect_lattice}). Therefore, although this ground state is neutral, in the sense that it is composed by $T_1$ vertices only, it should exhibit, in principle, a pair of opposite magnetic charges separated by a distance of the order of the lattice spacing. To better understand this picture one may think that an augmentation in the magnitude of the spin, for example, was caused by the inclusion of another (smaller) spin in the vertex, which is located at the same place and 
that points in the same direction as the increased one. In this case one has five spins instead of four at the adjacent vertices shared by this island, and thus there is no way to achieve neutrality in the vertex that contains the defect. Since vertices that do not satisfy the ice rule are viewed as magnetic monopoles, these defective vertices can also be viewed as a pair of monopoles. One may also easily arrive at the same conclusion by using a dumbbell picture as used by Castelnovo et al.~\cite{Castelnovo}. There is thus a pair of magnetic charges of magnitude $Q_D$, whose value depends on the unbalance at the vertices shared by the defective island.

\section{RESULTS}

In order to verify these assumptions, we consider now an elementary excitation in the system with one defect. It is a single pair of Nambu monopoles and its associated string placed in the vicinity of the static lattice defect as illustrated in Fig.~\ref{fig:defect_lattice}. We have analyzed the two particular string shapes shown in Fig.~\ref{fig:defect_lattice}; other string shapes were also studied, giving similar results. In Fig. \ref{fig:defect_lattice}, numbers $1$ and $2$ indicate the positions of the Nambu monopoles, with charges $-Q_{M}$ and $Q_{M}$ respectively, while numbers $3$ and $4$ indicate the extremes of the lattice defect, which are represented by a small yellow (light gray) arrow (at these points, as discussed above, two hypothetical opposite magnetic charges $-Q_{D}$ and $Q_{D}$ are positioned). Also, the two string shapes considered are referred to as strings $I$ and $II$ as shown in Fig.~\ref{fig:defect_lattice} in $(a)$ and $(b)$ respectively. In our calculations, the Nambu pair size 
$R$ (the 
smallest distance of separation between the two charges) is varied but only the position of charge $1$ will be shifted for convenience; position $2$ is kept fixed while positions $3$ and $4$ cannot change naturally, since we are considering a static defect. Making a suitable choice of the origin at position $2$, we get $|\vec{r}_1|=R$. Then, in principle, we now have four poles in the array: two coming from the static lattice defect and two from the induced excitation (Nambu pair).

Firstly, we would like to know the effects of this defect on the interaction potential between the monopoles $1$ and $2$, $V_{D}(R)$. The potential $V_D(R)$ can be obtained by calculating the system's energy for each configuration and subtracting the ground-state energy (see Refs.~\onlinecite{Mol1,Mol2}). To extract the effects of the defective island on the interaction energy we look for the difference $\Delta = V_{D}(R)-V_{N}(R)$, where $V_D(R)$ is the potential obtained for the system with the defective island and $V_{N}(R)$ is the potential obtained for a uniform system, where $|\vec{S}_l|=1$. Since $V_{N}(R)$ contains the interactions between the monopoles $1$ and $2$ and the string energy, $\Delta$ gives the interaction between the defective island and the monopoles $1$ and $2$ as well as the interaction between the defective island and the string. In this way we can write a general analytic expression for $\Delta$ by considering the Coulomb interaction energy between four charges ($1$ and $2$ with 
magnitude $Q_M$ and $3$ and $4$ with magnitude $Q_D$) added with the interaction between the defect and the string. The interaction between charges and strings and also between strings may be very complicated to explicitly write. Thus, for the moment, we are including in the general expression for $\Delta$ an \emph{ad hoc} term, such that $\Delta$ reads:
	\begin{equation}\label{e_interaction}
		\begin{array}{ccl}
		 	\Delta &=& K_{1}\left[ \dfrac{1}{|\vec{r}_{13}|}- \dfrac{1}{|\vec{r}_{14}|}+ \dfrac{1}{|\vec{r}_{24}|}- \dfrac{1}{|\vec{r}_{23}|} \right] \\
		 	\\
		 	& &+K_{2} \theta \left(\vec{R}_d\cdot\dfrac{\vec{r}_1}{|\vec{r}_1|}-|\vec{r}_1|\right) \, ,
		\end{array}
	\end{equation}
where
	\begin{equation}\label{k1}
		K_{1}= \dfrac{\mu_{0}}{4 \pi }Q_{D}Q_{M} \,
	\end{equation}
and $K_2$ are constants that must be determined, $\theta(z)$ is the step function ($\theta(z) =0$ for $z<0$ and $\theta(z)=1$ for $z>0$), $\vec{r}_{ij}$ is the distance between vertices $i$ and $j$, $\vec{r}_1$ is the position of charge $1$ and $\vec{R}_d$ is the position of the defective island. The first term of equation~\ref{e_interaction} is simply the Coulomb interaction energy between the four charges (the interactions between charges $1$ and $2$ are not present in $\Delta$ as well as the interaction between the defects $3$ and $4$). The second term represents the \emph{ad hoc} interaction of the string with the defect and will be discussed later.

\begin{figure}[hbt]
		\center
		\subfigure[String $I$.]{
			\includegraphics[width=7.0cm]{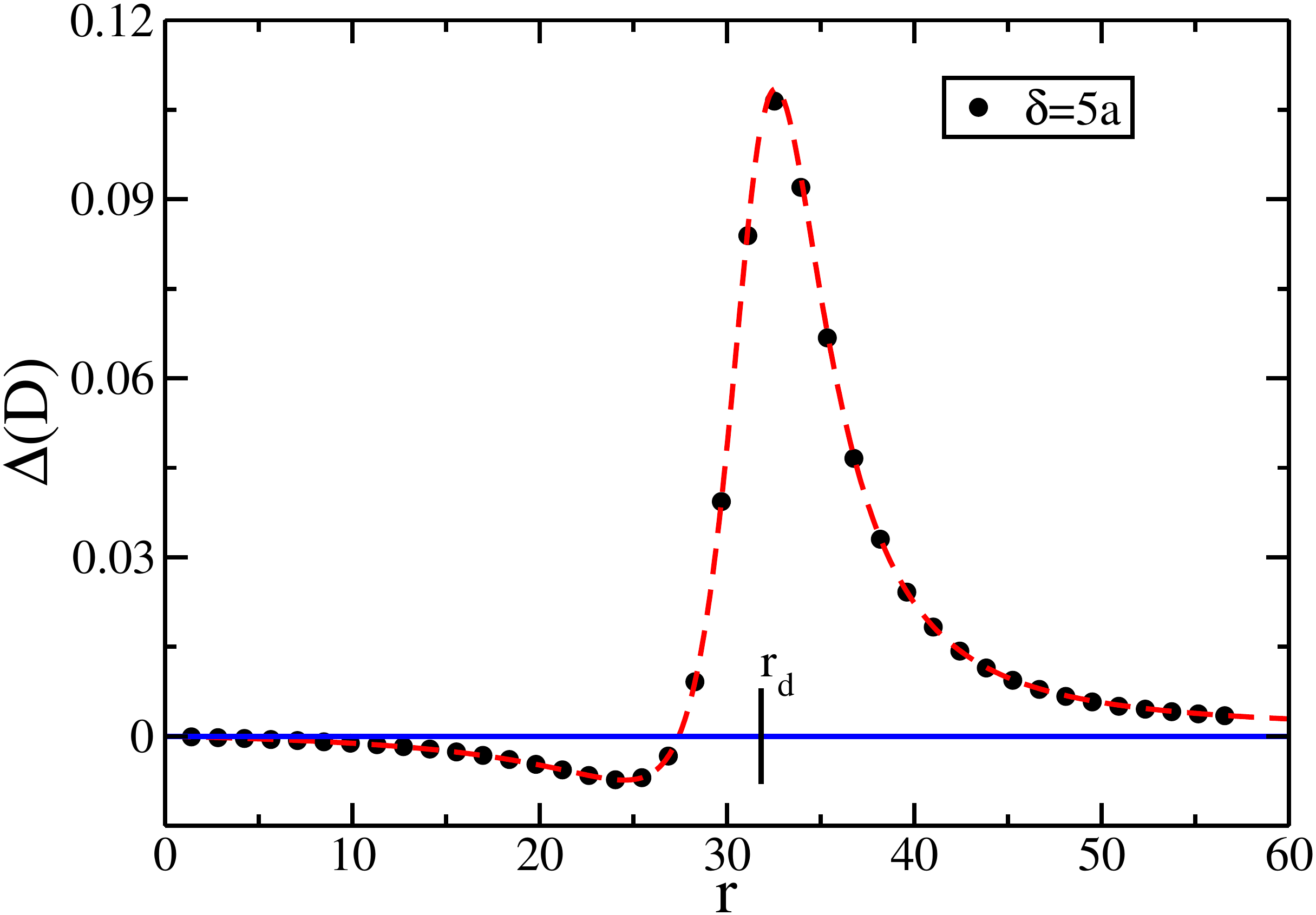}
			\label{fig:pot_string15}}
		\qquad
		\subfigure[String $II$.]{
			\includegraphics[width=7.0cm]{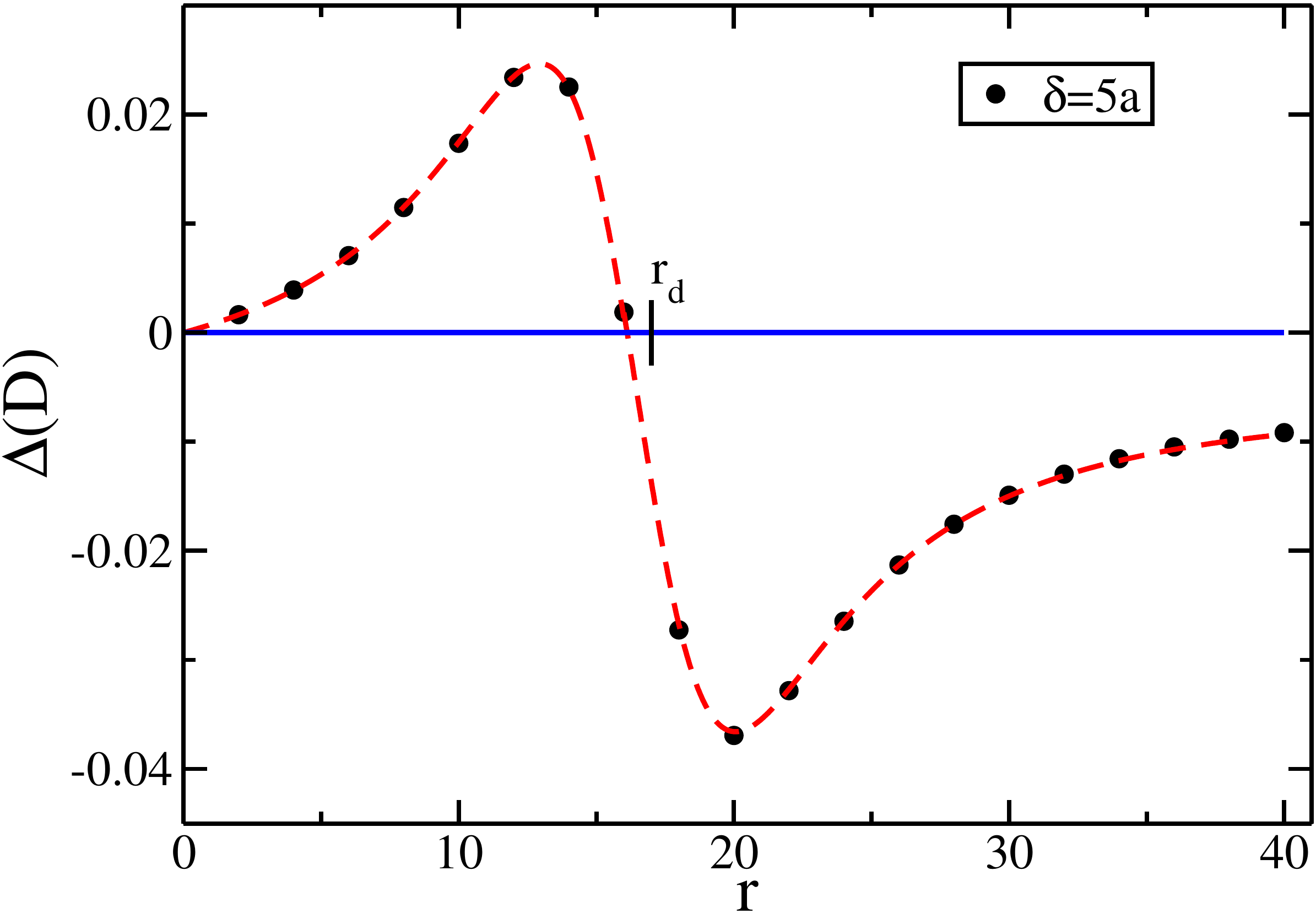}
			\label{fig:pot_string25}}
		\caption{(Color online) Data for $\Delta$ as a function of $r=R/a$, considering string type $I$ (a) and $II$ (b) (the string shapes $I$ and $II$ are shown in Fig.~\ref{fig:defect_lattice}). In these figures the smallest distance between position 1 and the defect, $\delta$, is larger than one spacing lattice. The magnitude of dipole moment is $|\vec{S}_{l}|=0$. The simulation data, $\Delta$, are the points and represent the case with $\delta=5a$. The dashed lines are the fits to expression \ref{e_interaction}.}
		\label{fig:pot_string5a}
	\end{figure}

Fig.~\ref{fig:pot_string5a} shows the potential $\Delta$ as a function of the distance between the Nambu monopoles $1$ and $2$ ($r=R/a$) for strings $I$ and $II$, using $|\vec{S}_{l}|=0$, i.e., considering a missing island in the system. The results presented here are for a lattice with size equal to $80a \times 80a$ (with 12,800 spins); however, several lattice sizes ($10a \leq L \leq 80a$) were also studied but not shown here since the results are almost the same. In this figure the smallest distance between the defect and the string is $\delta=5a$ ($\delta$ is measured as the distance between the line that connects the monopoles $1$ and $2$ and the defective island; note however that for the string shapes used here this distance is exactly the smallest distance between the string and the defective island). Since $\delta$ is relatively large we may consider that the defect does not effectively interact with the string, so that the constant $K_2$ may be set to zero.
 The dashed red line in Fig.~\ref{fig:pot_string5a} is a nonlinear curve fitting made by using Eq.(\ref{e_interaction}) with $K_2=0$. It can be seen that the Coulomb interaction between the Nambu monopoles (charges $1$ and $2$) and the defect charges ($3$ and $4$) correctly describes the data. Similar results are obtained for $0 \leq |\vec{S}_{l}| \leq 2$ and for any value of $\delta \geq 2a$.

These results show that a vacancy or even a defective island behave simply like a pair of opposite monopoles separated by a lattice spacing $a$ as suspected above. The maximum and minimum of the data in Fig.~\ref{fig:pot_string5a} can be understood by considering the repulsion and attraction between the mobile Nambu monopole $1$ and the defect charges $3$ and $4$. Indeed, the potential changes from repulsive to attractive, or vice-versa, as the monopole $1$ passes alongside the defect charges. The repulsion or attraction occurs if the monopole $1$ is closest to a defect charge of the same or opposite sign respectively. Another characteristic of this interaction concerns the presence of the string. Since $K_2$ was set to zero to fit the data, one could conclude that the string connecting the Nambu monopoles does not cause any effect on the interaction if its distance from the defect is relatively large. This is really the
  situation, as we will explain later.

	\begin{figure}[hbt]
		\centering
		\includegraphics[width=7.0cm]{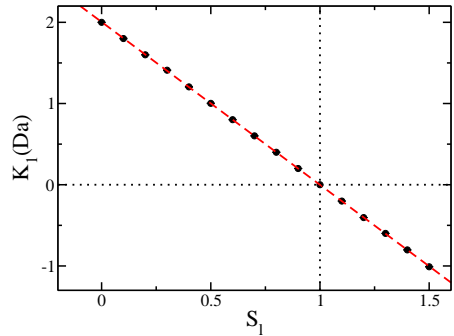}
		\caption{(Color online) Data for $K_{1}$ as a function of the size of the defective island. Observe the linear behavior of $K_{1}$ as $S_{l}$ is increased.}
		\label{fig:k1f}
	\end{figure}

\begin{figure}[hbt]
		\centering
		\includegraphics[width=8.0cm]{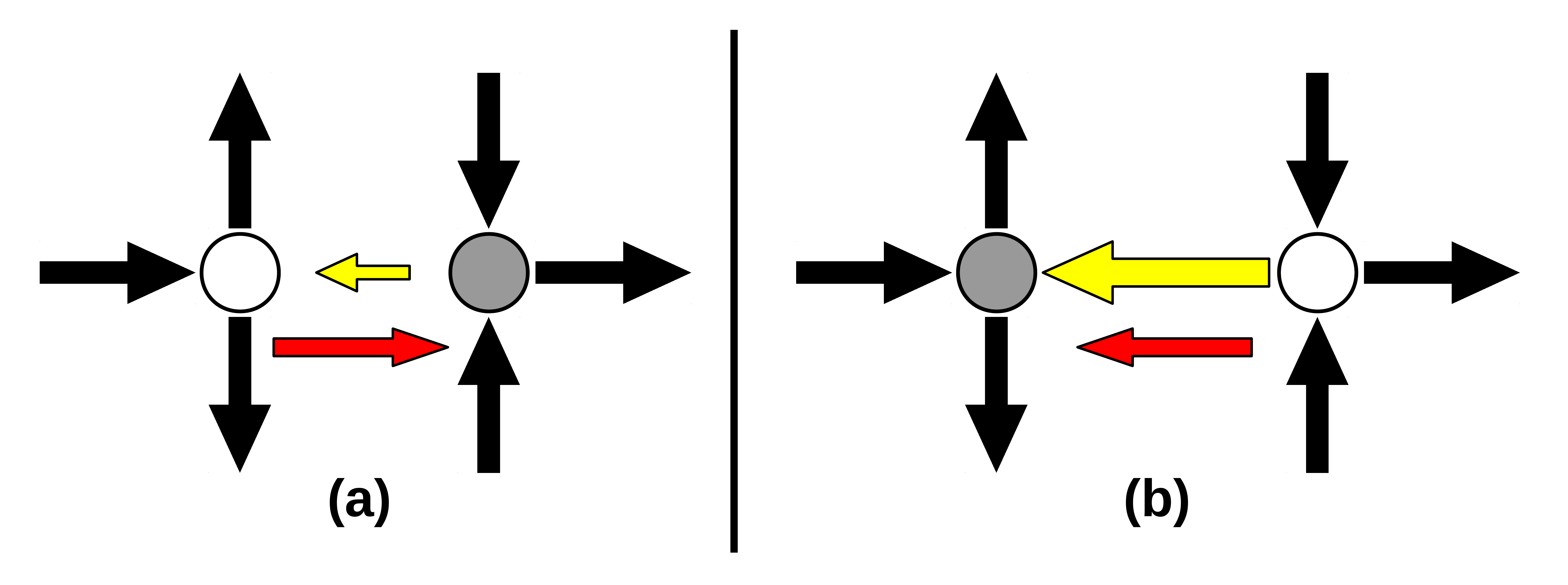}
		\caption{(Color online) Ground state configuration of the system around the defective island for (a) $S_{l}$ smaller (b) $S_{l}$ greater than other islands. The white circle represents the negative charge induced by the lattice defect while, the gray circle represents the positive charge.}
		\label{fig:cargas_induzidas}
	\end{figure}

Figure \ref{fig:k1f} shows the fitted values ($K_{1}$) as a function of $S_{l}$, obtained for type $I$ string (the same result was also obtained for type $II$ string). The red dashed line is a linear regression.
For $S_{l} = 0$, our results show that $K_{1} \approx 2 \, Da$. Besides, using the fact that $K_1$ is given in units of $Da$, it is easy to show that $Q_D=\dfrac{\mu}{a}\dfrac{K_1}{Q_M}$ and since $Q_{M}\approx 2$, $Q_D\approx 1 \approx Q_M/2$. It leads to $|Q_{D}| \approx |Q_{M}|/2$, which should be expected since the defect topology is an arrangement with configuration 2-in/1-out and vice-versa.
The magnitude of $K_{1}$ decreases with increasing $S_{l}$, vanishing, as expected, when $S_{l}=1$ (which is the case of a ``perfect array"). For $S_{l}>1$, the sign of $K_{1}$ changes, indicating that there is a switch in the position of the positive and negative charges produced by the defect, as shown in figure \ref{fig:cargas_induzidas}. In this figure, the white and gray circles represent the negative and positive charges induced by the lattice defect. In fact, the switch of the position of the induced charges can be easily seen by observing the change in the net magnetic moment (red (dark gray) arrow) on the vertices that form the lattice defect when $S_{l}$ is smaller or greater than the other islands spin. Therefore, the effect of varying $S_{l}$ from values smaller than $1$ to values larger than $1$ is the same as that of inverting the effective magnetic moment of an island from an arbitraty value $g$ to $2-g$ ($0 \leq g \leq 1$); hence, the characteristic effect 
of a vacancy is essentially the same as that caused by a defect island with a spin whose magnitude is twice ($S_{l}=2$ ) the magnitude of the spin of the normal islands (for defects with $S_{l}=0$ and $S_{l}=2$, the magnetic moment has the same modulus but it points in opposite directions).

\begin{figure}[hbt]
		\center
		\subfigure[String $I$.]{
			\includegraphics[width=7.0cm]{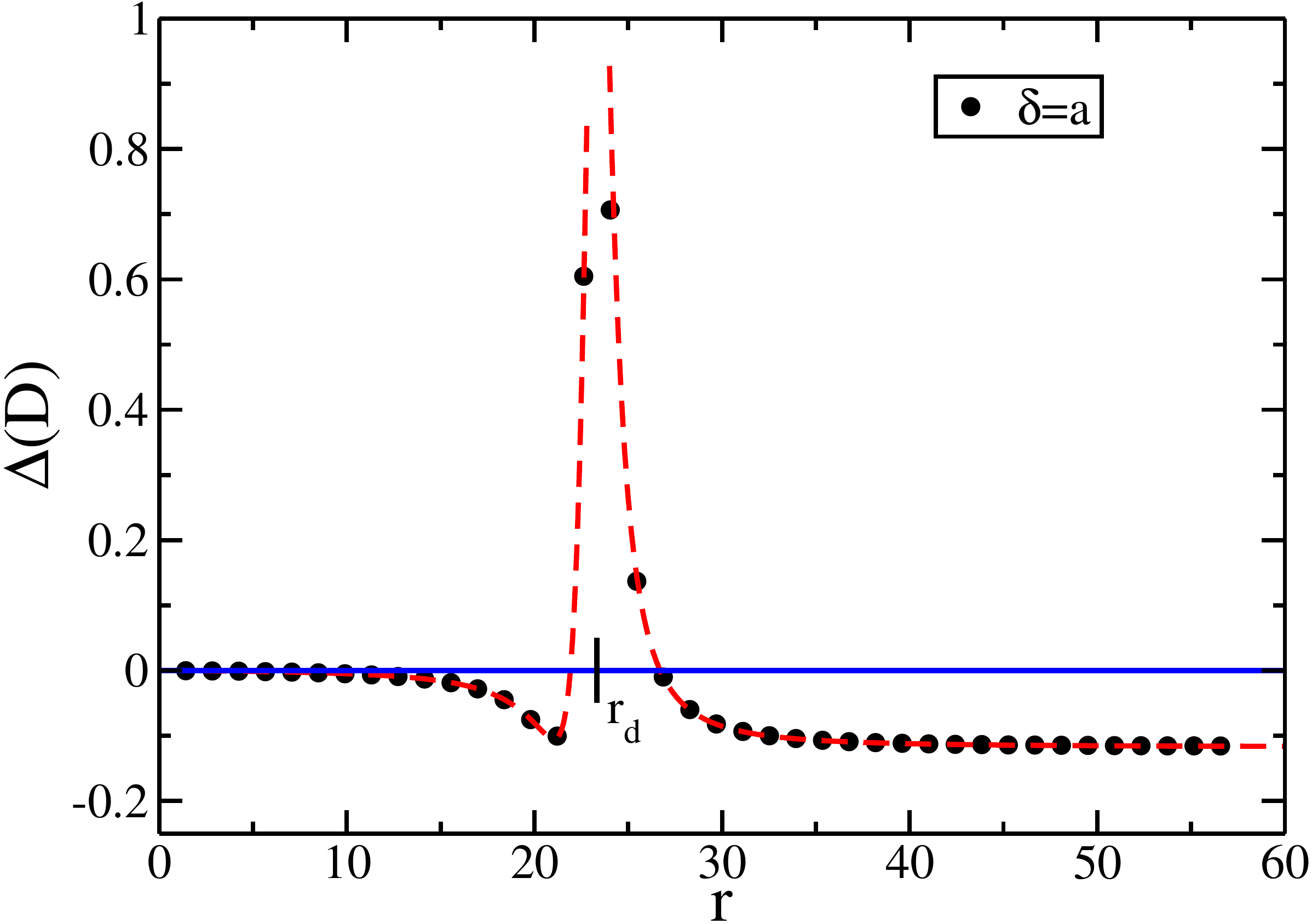}
			\label{fig:pot_string11}}
		\quad
		\subfigure[String $II$.]{
			\includegraphics[width=7.0cm]{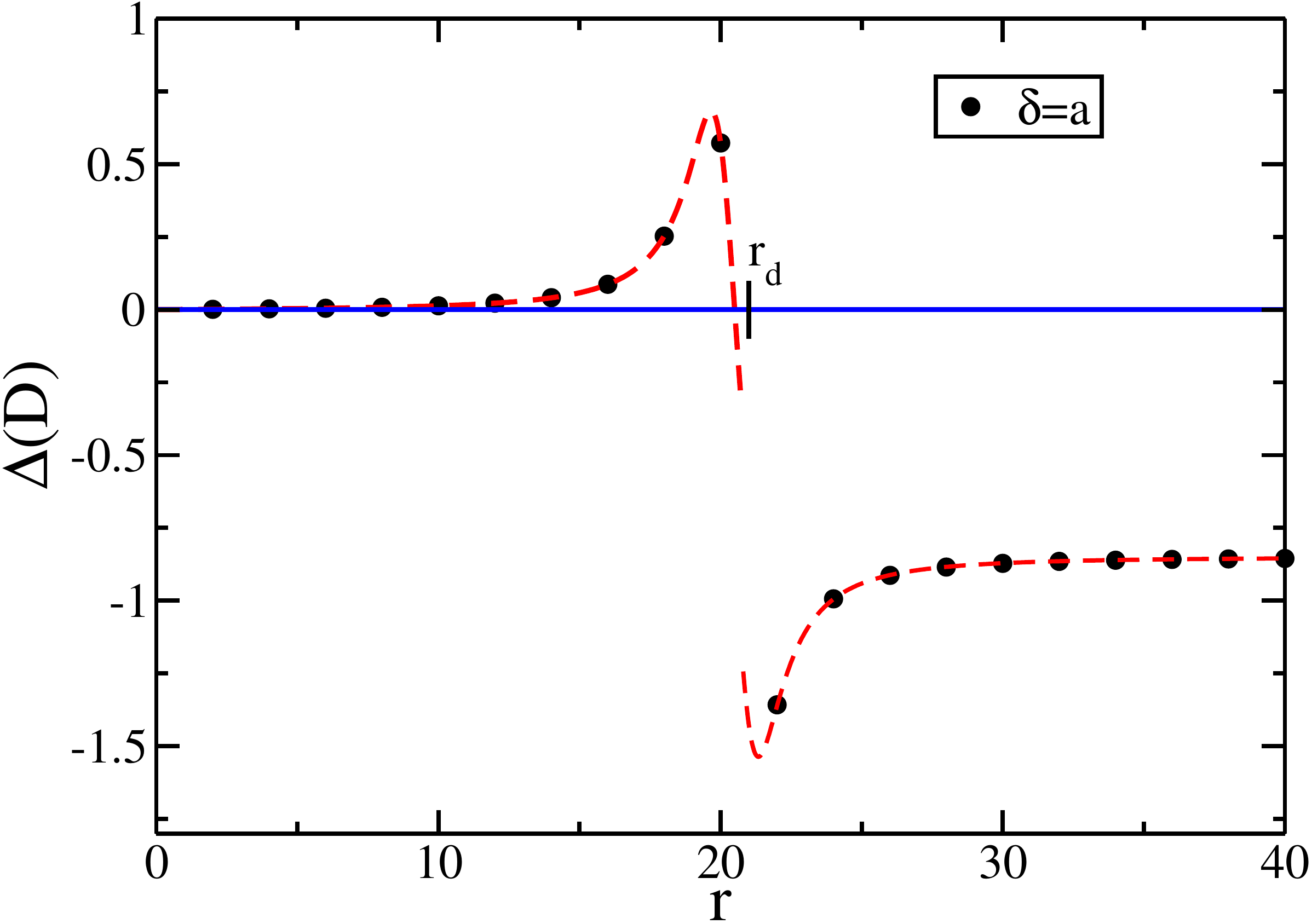}
			\label{fig:pot_string12}}
			\caption{(Color online) Data for $\Delta$ as a function of $r=R/a$, considering strings $I$ (a) and $II$ (b), for $\delta = 1a$ and $S_{l}=0$. The simulation data for $\Delta$ are the black points and the red dashed line is the fit to expression \ref{e_interaction}.}
		\label{fig:string1a}
	\end{figure}

\begin{figure}[hbt]
		\centering
		\includegraphics[width=8.0cm]{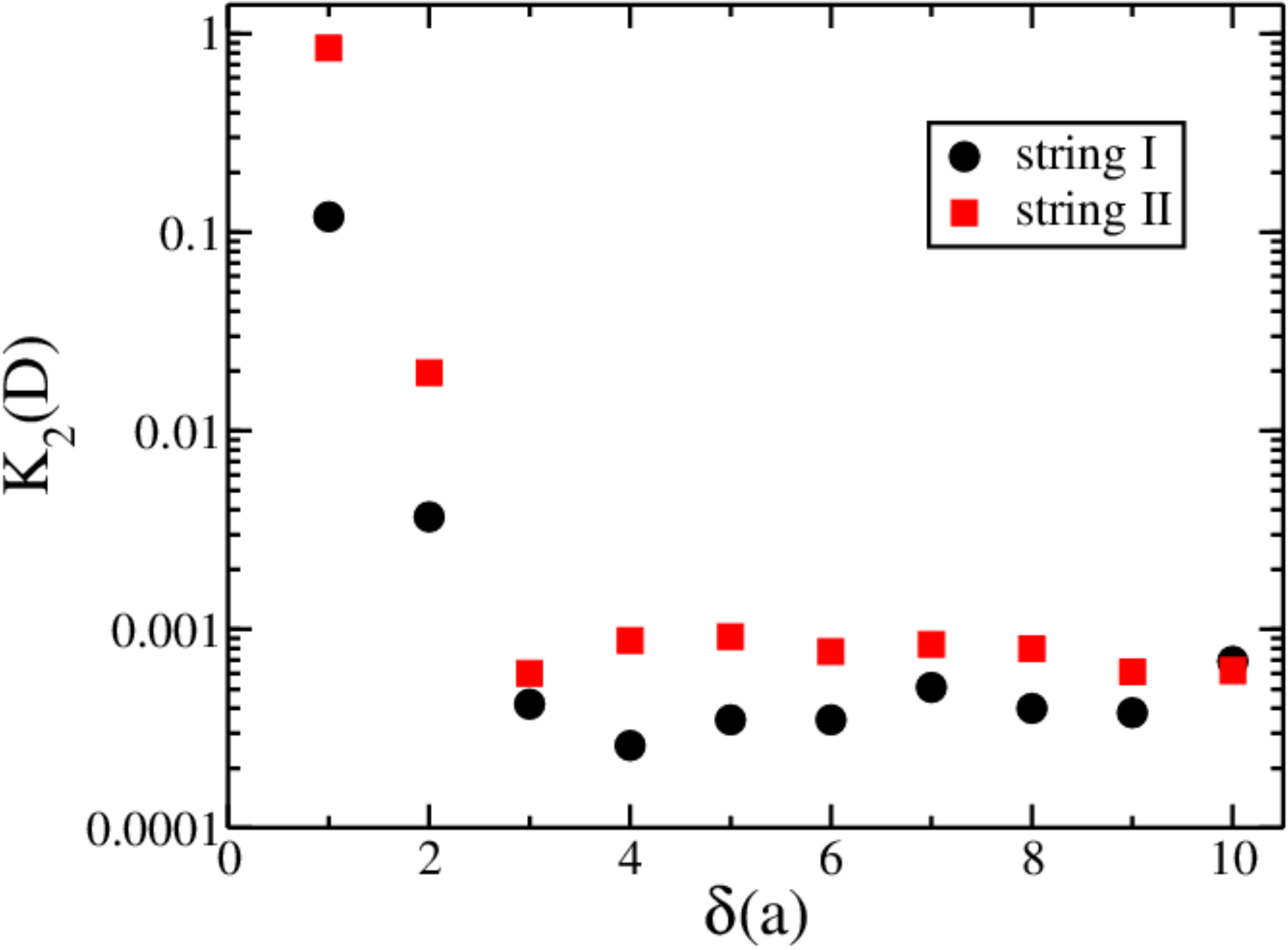}
		\caption{(Color online) The interaction constant between string and defect, $K_2$, as a function of their separation, $\delta$. The black circles and red squares are the fitting data obtained for the strings I and II respectively. }
		\label{fig:K2}
	\end{figure}

On the other hand, if the ``moving'' Nambu monopole $1$ passes close to the lattice defect at a distance smaller than $2a$ (on the order of $1a$), a substantial difference in the interaction potential can be noted, as shown in Fig. \ref{fig:string1a}, which is obtained for string shapes $I$ and $II$ near a vacancy ($\delta =1 a$). For large values of $r$ we can see that $\Delta$ goes to a constant value, while in Fig.~\ref{fig:pot_string5a} it goes to zero. This difference is attributed to the interaction between the string and defect, which as can be seen in Fig.~\ref{fig:pot_string5a} decays very quickly. Then, we may expect that the string interacts only with very close objects. In this way, we may see that when all parts of the string are far from the defect, there is no contribution from its interaction with the defect to the total energy. On the other hand, if a segment of the string is close enough to the defect (distance
  smaller than $2$ lattice spacings), only the small segment that is close enough to the defect interacts with it. This justifies the \emph{ad hoc} term included in Eq.~\ref{e_interaction}. When the ``distance'' between the defect and the ``moving'' monopole $\left(r_d-r_1\text{, where }r_d=\vec{R}_d\cdot\dfrac{\vec{r}_1}{|\vec{r}_1|}\right)$ is negative, the string has not crossed the
defect yet, and thus, it is not close enough to contribute to the total energy. On the other hand, for $r_d >0$ there is a segment of the string at a distance $\delta$ from the defect and for $\delta < 2a$ this segment contributes with a constant value $K_2$ to the total energy. In Fig.~\ref{fig:string1a}, the dashed red line was obtained by doing a nonlinear curve fitting according to general Eq.~\ref{e_interaction}, in such a way that for $r<r_d$, $K_2$ was set to zero and then, keeping $K_{1}$ fixed the remaining points ($r > r_d$) were fitted for arbitrary $K_2$.  In Fig.~\ref{fig:K2} we show the results for the constant $K_2$ as a function of $\delta$. As can be seen, $K_2$ has a significant value only for $\delta<3$. The fact that the main interaction between defects is, in general, short ranged is in agreement with previous results from Ref.~\onlinecite{Morgan}, where the energetics of excitations above a thermalized state are studied.

\section{Magnetic field lines}

\begin{figure}[hbt]
              \includegraphics[width= \linewidth]{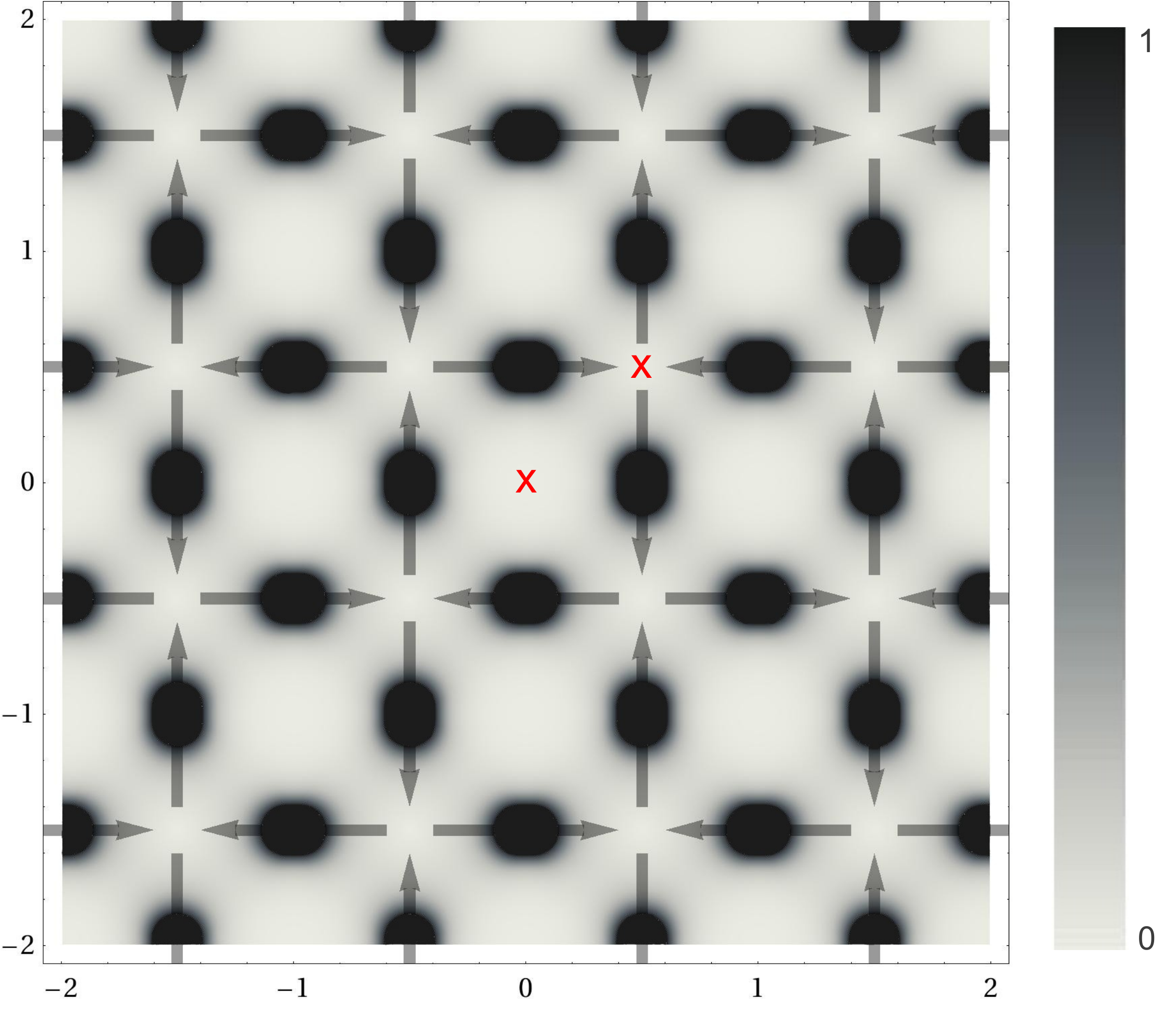}
			\caption{(Color online) Magnetic field intensity of the system's ground state. At the center of the plaquettes and at the centers of the vertices, indicated by the red crosses, the field goes to zero (white regions). These points were used to obtain the magnetic field produced by the excitations alone. At the right side of the figure the color palette for the magnetic field intensity is shown in normalized units.}
		\label{fig:intensity}
	\end{figure}

\begin{figure}[hbt]
			\includegraphics[width = \linewidth]{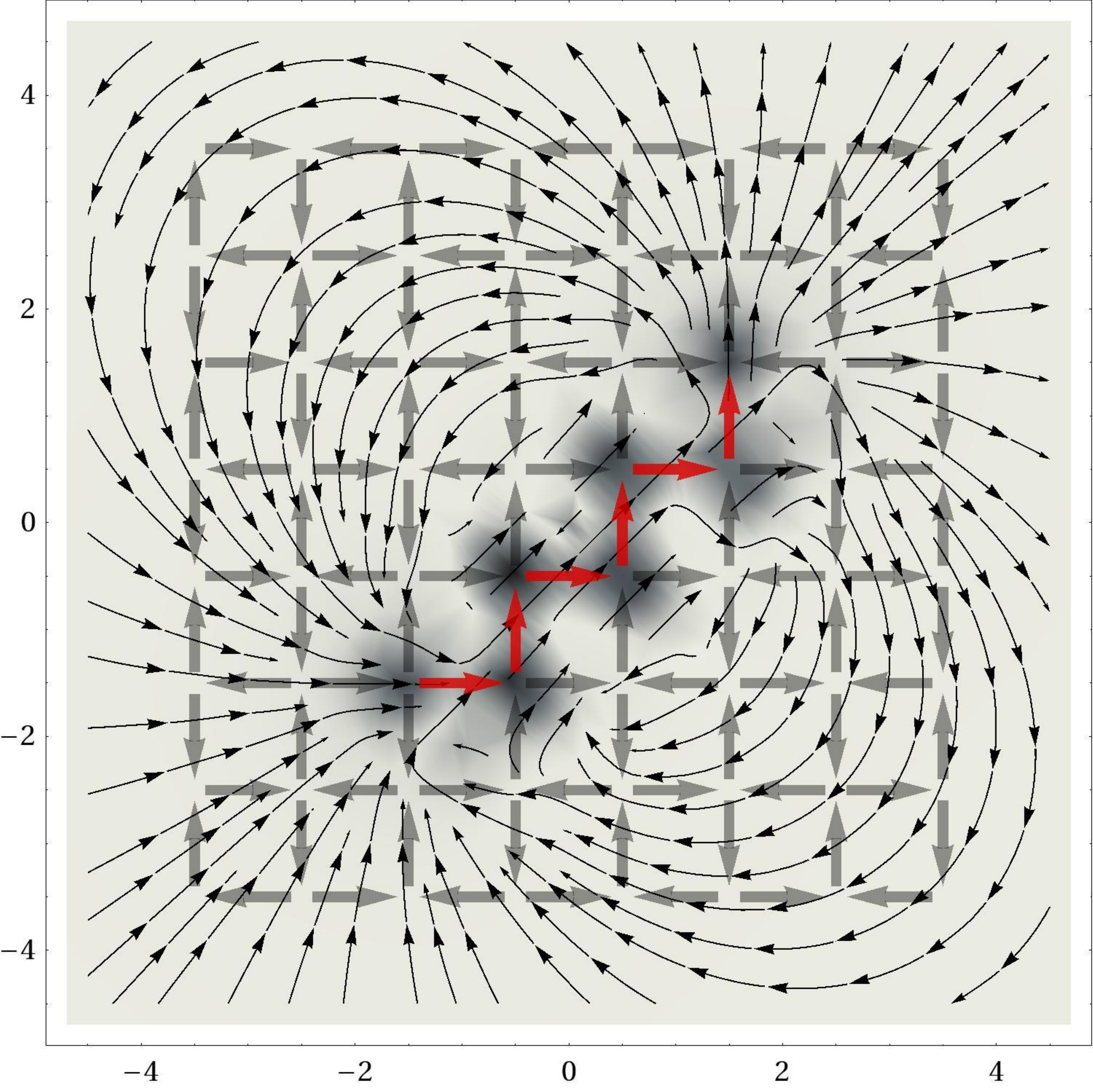}
			\caption{(Color online) This figure exhibits the magnetic field lines of a pair of Nambu monopoles and its string (the spins located between the dark gray vertices or red spins) considering the field produced by the excitation alone. Only a small portion of the system is shown for clarity.}
		\label{fig:lines1}
	\end{figure}

\begin{figure*}[t]
		\center
		\subfigure[Electric charges.]{
			\includegraphics[width=0.45 \linewidth]{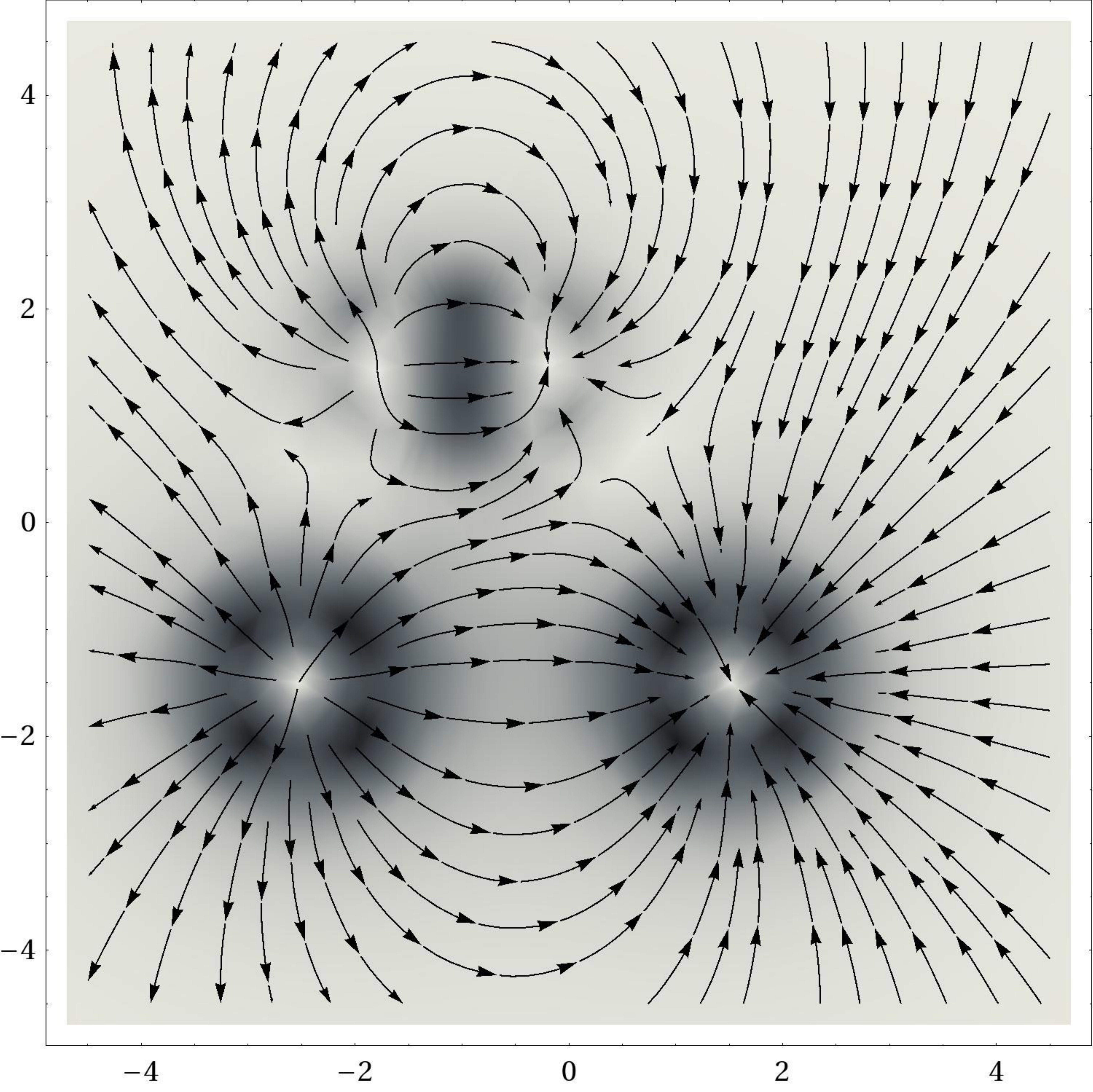}}
		\qquad
		\subfigure[Nambu monopoles connected by a type $II$ string and a vacancy.]{
			\includegraphics[width=0.45 \linewidth]{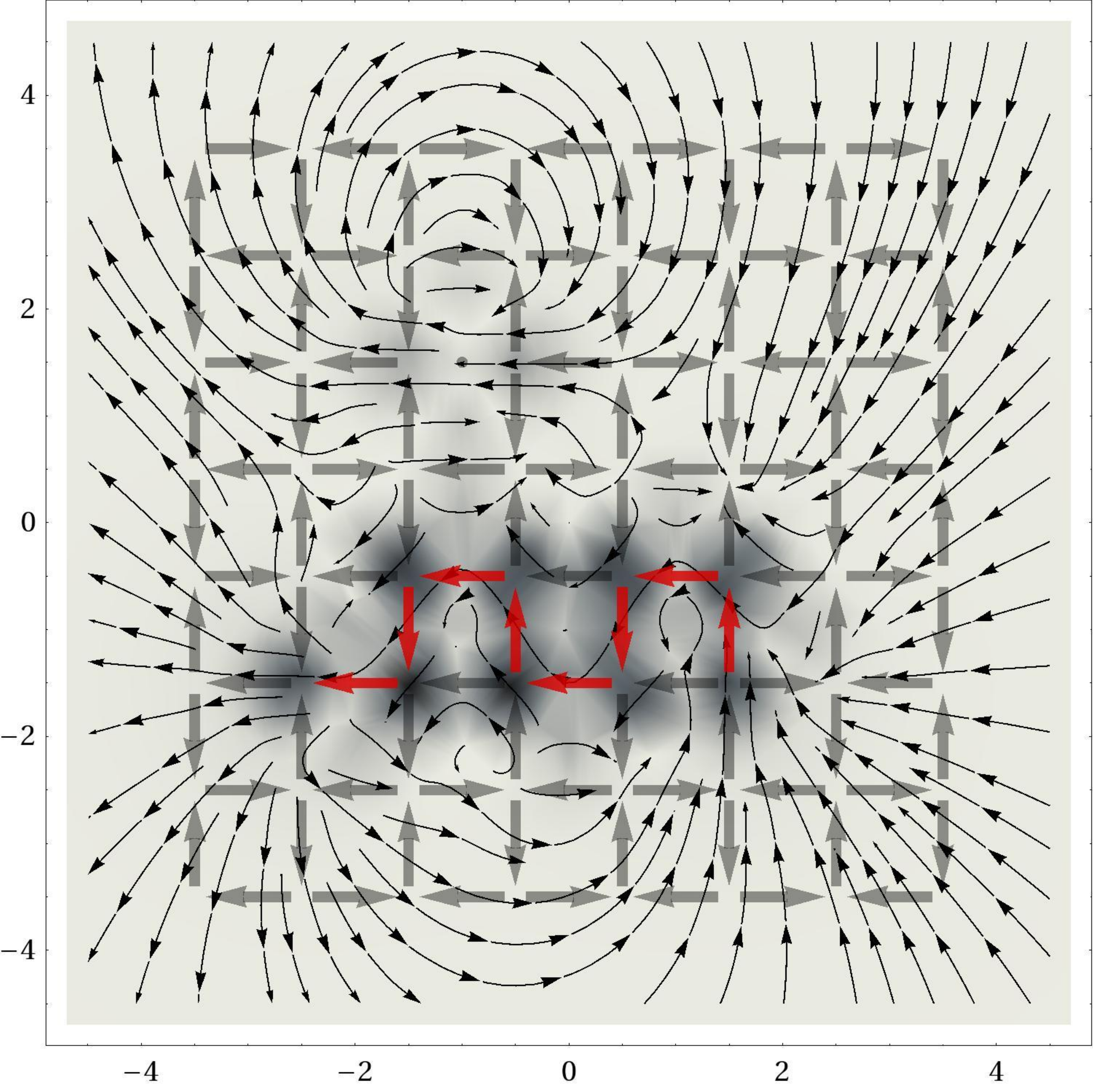}}
			\caption{(Color online) (a) Electric field lines produced by four electric charges: two of unity magnitude representing the Nambu monopoles and two of one half magnitude representing the vacancy's charges. (b) Magnetic field lines of the spin configuration obtained by considering the field produced by the excitation alone. The red spins (the spins located between the dark gray vertices) were flipped to produce the Nambu monopoles and its string. Only a small portion of the system is shown. The vacancy is placed approximately at (-1,2).
		\label{fig:lines2}}
	\end{figure*}

The above results give strong support to the monopole-like picture for the excitations and defects of the artificial square spin ice as presented in Ref.~\onlinecite{Mol1}. To give further support to this scenario we have also analyzed the magnetic field lines for this configuration of charges. We start our analysis by presenting in Fig.~\ref{fig:intensity} a color map of the magnetic field intensity of a perfect system in its ground-state, where all islands have the same spin value. It is easy to see that the field is null at the centers of the plaquettes as well as at the centers of the vertices (two of these points are indicated by red crosses in the figure). This fact allows us to obtain the field produced by the excitations alone without considering the detailed structure of the magnetic field produced by the dipoles. Thus, in our calculation, the magnetic field produced by the excitation can be obtained by simply inverting spins (creating excitations) and then calculating the resulting magnetic field 
at the center of the plaquettes and at the center of the vertices, i.e., by calculating the magnetic field at the points where it is zero in the ground-state. In Fig.~\ref{fig:lines1}, we show the stream lines of the aforementioned field for a configuration where the red spins were flipped. We notice that the magnetic field lines far from the flipped spins are very similar to the field lines of a pair of electric charges, while in the space between them the magnetic field follows the string. It becomes clear then that the string carries the magnetic flux back from one charge to the other. In Fig.~\ref{fig:lines2} (b) we show the stream lines of a configuration containing a vacancy and a pair of magnetic monopoles and its associated string while, for effect of comparison, in Fig.~\ref{fig:lines2} (a) we show the electric field produced by two unity charges located at the same position of the Nambu monopoles and a pair of one half charges located at the same position of the vertices shared by the defective 
island. Apart from the region where the string is present, the similarities between these two figures is remarkable. Although very simple, this analysis gives further evidences for the monopole-like behavior of excitations and defects in the artificial square spin ice.

\section{Conclusion and prospects}

We have studied the interaction of two magnetic monopoles (and the energetic string connecting them) with a lattice defect present in the square spin ice array. We notice an interesting resemblance between the single defect and a static pair of monopoles separated by one lattice spacing. The strength of the magnetic charges of this small defect was obtained as a function of the magnetic moment of the defective island (Fig.~\ref{fig:k1f}). Defects with $S_{l}=0$ (vacancy) and $S_{l}=2$ (double spin) produce similar effects in the lattice, since they have the same magnetic charges (placed in opposite positions). There is also a short range interaction between the string and the lattice defect, which can be attractive or repulsive, depending on the orientation and local shape of the string. Our results are an important step towards understanding how lattice defects could change the thermodynamics of artificial spin ices\cite{Silva}. For instance, considering an array with a finite density $\varrho$ of defects 
as done, for example, in Ref.~\onlinecite{Budrikis}, it should be important to know how the properties of the system change as $\varrho$ increases and how defects could affect the formation of the ground state experimentally (a problem usually found in experiments with artificial square ices\cite{Wang,Morgan}). In general, we expect that the presence of a finite density of these lattice defects will strongly distort the path of the strings and they could even, break or join some different strings. In addition, since the defects act as small pairs of charges, we expect that the presence of them in the lattice should affect the monopole average separation and density (as calculated in Ref.~\onlinecite{Silva}) and, as $\varrho$ increases,  the peak position of these quantities should be altered for lower temperatures.  Of course, it may also have an effect on the entropically driven monopole unbinding~\cite{Silva} and the critical temperature may decrease as $\varrho$ increases;  probably, the 
fact that the 
effect of the defect on the Nambu strings seems irrelevant when their distance exceeds a few lattice constants means that the transition is unaffected by sparse disorder, but there might be a critical density of islands above which the entropic oscillation of the strings can get pinned thus destroying the transition. A more detailed study of these questions is currently in progress.

A simple way to model unintentional defects in the system is to suppose that the islands have a Gaussian size distribution around a mean value. In a model of point dipoles this would be achieved by considering a Gaussian distribution of the spins' magnitudes. In this case, one may expect that, for a small variance of the size distribution, the ground state would be the same as in the perfect system. However, for a large variance or for a system where the defects are not randomly distributed, we may expect some differences in the ground state. For instance, one may expect the formation of an ordered arrangement of charges (like a crystal of charges) similar to what happens in a kagome lattice~\cite{Chern11} and in a rectangular lattice~\cite{Nascimento12}. The control of some defects (for example, inducing stronger or smaller variances of the size distribution) may thus be used to facilitate the experimental achievement of the system's ground-state.

Another interesting point is the possibility to construct tailor designed systems to achieve some desired property. Since the presence of a defective island can be interpreted in terms of the induced charges at the vertices shared by it, one can think of designing, for example, a magnetic capacitor-like system. This would be constructed by designing a system where all spins in a stripe immersed in a square system have islands smaller than all others, for example, in such a way that, in the edges of this stripe, there will be residual charges as far as an ice-like state is achieved. The presence of these residual charges may significantly change the behavior of other excitations inside this capacitor. A more detailed analysis of this hypothesis is under consideration.

\section*{Acknowledgements}
The authors thank CNPq, FAPEMIG, CAPES and FUNARBE (Brazilian agencies) for financial support.



\end{document}